\documentclass[twocolumn,showpacs,preprintnumbers,amsmath,amssymb]{revtex4-1}

\usepackage{graphicx}
\usepackage{dcolumn}
\usepackage{bm}

\begin{document}

\preprint{APS/123-QED}

\title{Probability distribution function for a solid with vacancies}

\author{Leonid S. Metlov}

 \email{lsmet@fti.dn.ua}
\affiliation{Donetsk Institute of Physics and Engineering, Ukrainian
Academy of Sciences,
\\83114, R.Luxemburg str. 72, Donetsk, Ukraine
}%

\date{\today}

\begin{abstract}
Expression for probability distribution is got taking into account a presence and removal of degeneracy on the microstates. 
Its application allows to describe the process of melting of solids, as saltatory phase transition of the first kind 
without bringing in of concept of the order parameter. 
\end{abstract}

\maketitle


For the case of a solid with vacancies the expressions for the thermodynamic probability W and the configuration entropy 
$S_{c}$ in supposition of the total degeneracy of the microstates had formerly got \cite{f55} 
  \begin{equation}\label{b1}
W=\dfrac{(N+n)!}{N!n!},
  \end{equation}
  \begin{equation}\label{b2}
S_{c}=k_{B}\ln W,
  \end{equation}
where $N$ is a total number of atoms in a crystal, $n$ is a number of vacancies, $k_{B}$ is the Boltzmann constant.

The total probability of the state with n vacancies contains a limiting exponential multiplier, probability of Gibbs 
for this microscopic state \cite{f55, s89, g97}
  \begin{equation}\label{b3}
f(n)=\dfrac{W}{Z}\exp(-\dfrac{U(n)}{k_{B}T}),
  \end{equation}
where $Z$ is the normalized statistical sum, $U(n)$ is the internal energy taking into account existence of vacancies, 
$T$ is the temperature. In the case of total degeneracy the internal energy is the linear function of number of 
vacancies \cite{f55}
  \begin{equation}\label{b4}
U(n)=U_{0}+u_{0}n,
  \end{equation}
where $U_{0}$ is the internal energy without taking in account of vacancy contribution, $u_{0}$ is the energy of a vacancy 
which in this case is a constant for all possible configurations of atoms of a system.

The condition of total degeneracy is executed more or less exactly for the low concentration (number) of vacancies, 
when it is possible to ignore the interaction between vacancies. At the same time, this condition is not executed 
for those configurations for which vacancies are located close to each other or merges at all (bi-vacancies, triple 
vacancies, vacancy pores) and at the high concentration or density of vacancies. In these cases it is necessary to take 
into account the removal of condition of total degeneracy due to the interaction between vacancies. 
As it is known, coupling interaction reduces the total energy of a system that reduces the effect of action of limiting 
exponential multiplier in Eq. \ref{b3} and resulting to possibility of appearance of long-living configurations, which, thus, 
can resist more numerous configurations with more high energy.

The probability distribution function (PDF) with total (internal) energy $E_{l}$ equals \cite{p71}
  \begin{equation}\label{b5}
f(n)=\dfrac{w(E_{l})}{Z}\exp(-\dfrac{E_{l}}{k_{B}T}),
  \end{equation}
where $w(E_{l})$ is distribution of states with energy or number of the microstates (configurations) with energy of $E_{l}$, 
$Z$ is statsum over all energetic states of the system
  \begin{equation}\label{b6}
Z=\sum_{l=1}^{All states} w(E_{l})\exp(-\dfrac{E_{l}}{k_{B}T}),
  \end{equation}

The states are numbered in order of its energy growth of $E_{l+1} > E_{l}$. Distribution $w(E_{l})$ depends of the number 
of vacancies and of symmetry of their location (ordering). On this account, it is impossible to fix one-to-one 
correspondence between internal energy and number or concentration of vacancies. Finding of function $w(E_{l})$ is 
an intricate combination problem, determination of power spectrum of $E_{l}$, as the set of acceptable energies for 
$l$ state, is also a stubborn problem. However, such one-to-one correspondence can be got for the average value of 
internal energy $U$ on all of the states at the fixed number of vacancies $n$. Using expression for PDF (\ref{b5}), 
it is possible to write down (see closed idea in \cite{p71}, p. 147)
  \begin{equation}\label{b7}
U(n)=\dfrac{1}{Z}\sum_{l=1}^{All states (n)} E_{l} w(E_{l})\exp(-\dfrac{E_{l}}{k_{B}T}),
  \end{equation}
Here unlike a case of Eqs. \ref{b1} - \ref{b4} the internal energy is not a linear function, but is the function of 
general view. The increase of fraction of the symmetric low-energy states diminishes the average value of the internal energy. 
The transition to the states, at which vacancies at the high degree of their concentration can pass to higher-energy extended 
(and more mobile) states \cite{gfs80, gfs83, tt06}, results in growth of the internal energy. For the reflection of all of these 
properties of the internal energy it presents in a form of polynomial with alternating signs
  \begin{equation}\label{b8}
U=U_{0}+ \sum_{k=1}^{K} \frac{(-1)^{k-1}}{k}u_{k-1}n^k   ,
  \end{equation}

Ignoring degeneracy within of one set of the states with the same number of vacancies $n$, for such average 
internal energy we can write down the «effective» function of distribution of probabilities in a form Eq. \ref{b3} 
with determinations (\ref{b1}), (\ref{b2}) and (\ref{b8}). Using such PDF one can describe processes of melting of solids, 
as phase transitions of the first kind \cite{m11}.

Values $n$, at which PDF has extreme values (equilibrium states), can be found from the follow transcendent equation \cite{m11}
  \begin{equation}\label{b9}
\sum_{k=1}^{N+n}\frac{1}{k}-\sum_{k=1}^{n}\frac{1}{k}-\frac{u}{kT_{B}}=0.
  \end{equation}
where
  \begin{equation}\label{b10}
u\equiv \frac{\partial U(n)}{\partial n}= \sum_{k=0}^{K-1} (-1)^{k}u_{k}n^k 
  \end{equation}
is energy or chemical potential of a vacancy.

\begin{figure}
\hspace{0.06 cm}
\includegraphics [width=3 in] {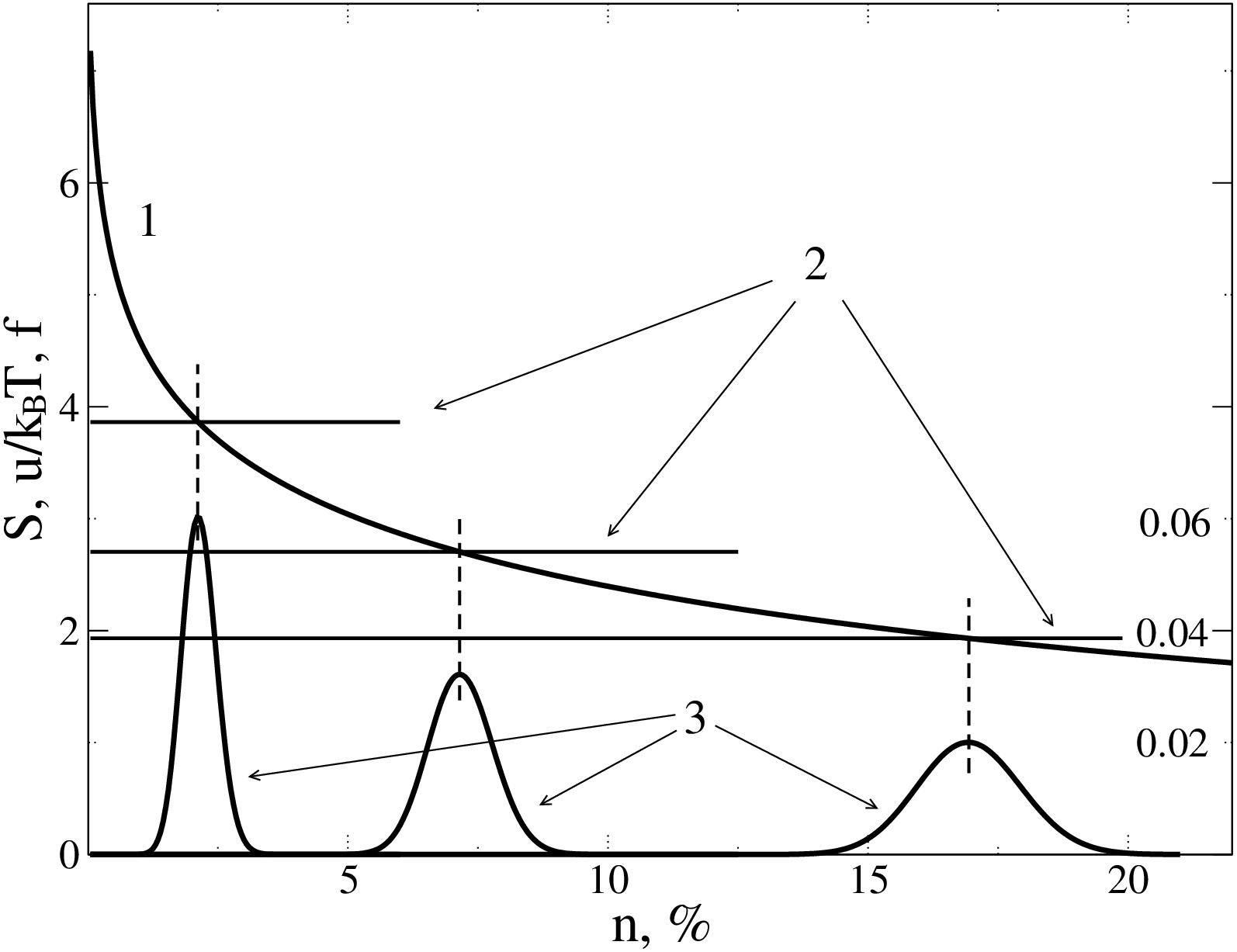}
\caption{\label{f1} Solution of Eq. \ref{b9} at constant energy of vacancy. The fundamental curve $S$ (1), 
the energy of vacancy $u$ (2) and PDF (3): $u_{0} = 0.1; 0.07; 0.05 (eV)$. Other parameters equals zero, $T=300K$. 
The right-hand scale is for PDF.}
\end{figure}

\begin{figure}
\hspace{0.06 cm}
\includegraphics [width=3 in] {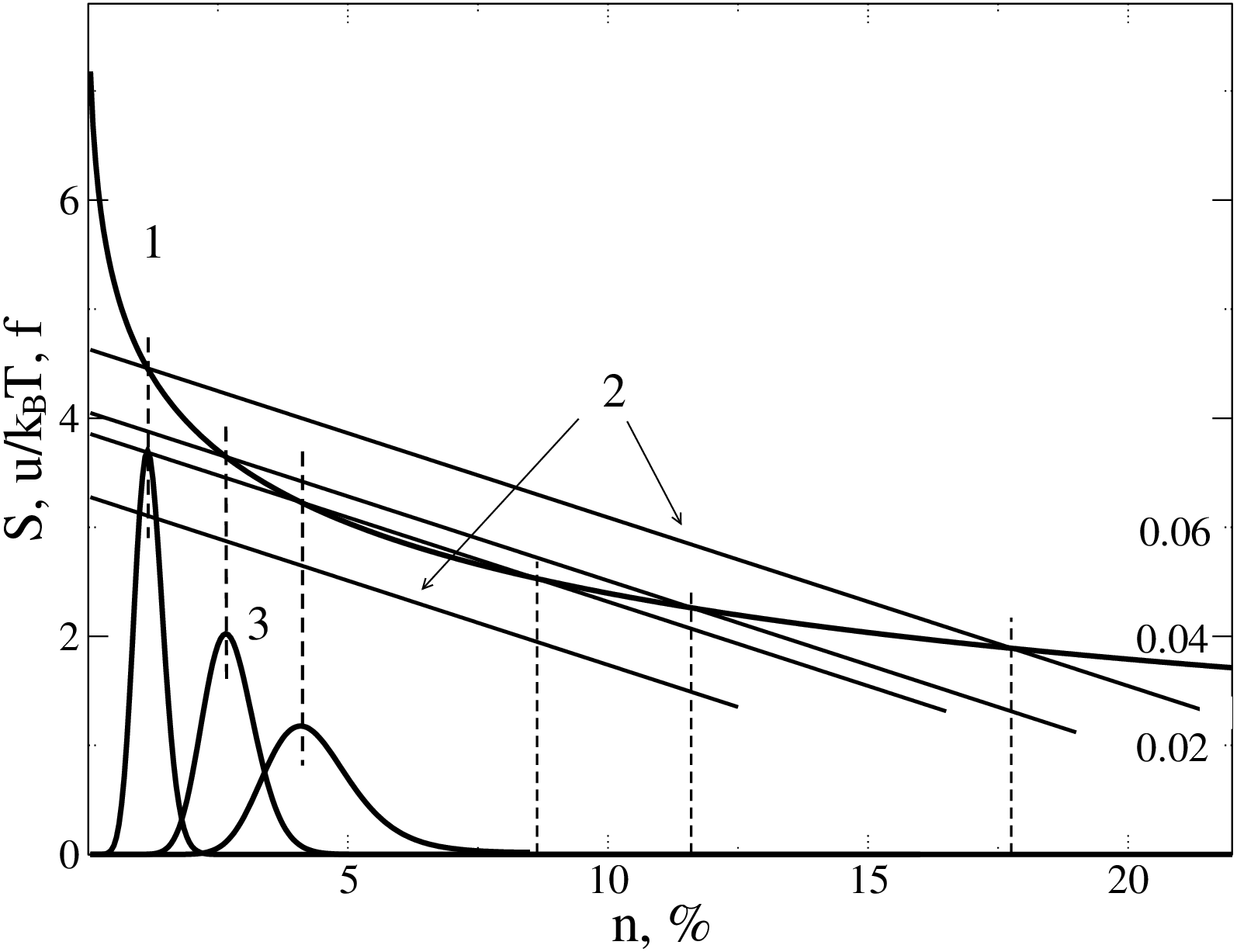}
\caption{\label{f2} Solution of Eq. \ref{b9} at linear dependence of vacancy energy of the vacancy concentration. 
The fundamental curve $S$ (1), the energy of vacancy $u$ (2) and PDF (3): $u_{0} = 0.12; 0.105; 0.1; 0.085 (eV)$, 
$u1 = 0.0002 (eV)$. Other parameters equals zero, $T=300K$. The right-hand scale is for PDF.}
\end{figure}

\begin{figure}
\hspace{0.06 cm}
\includegraphics [width=3 in] {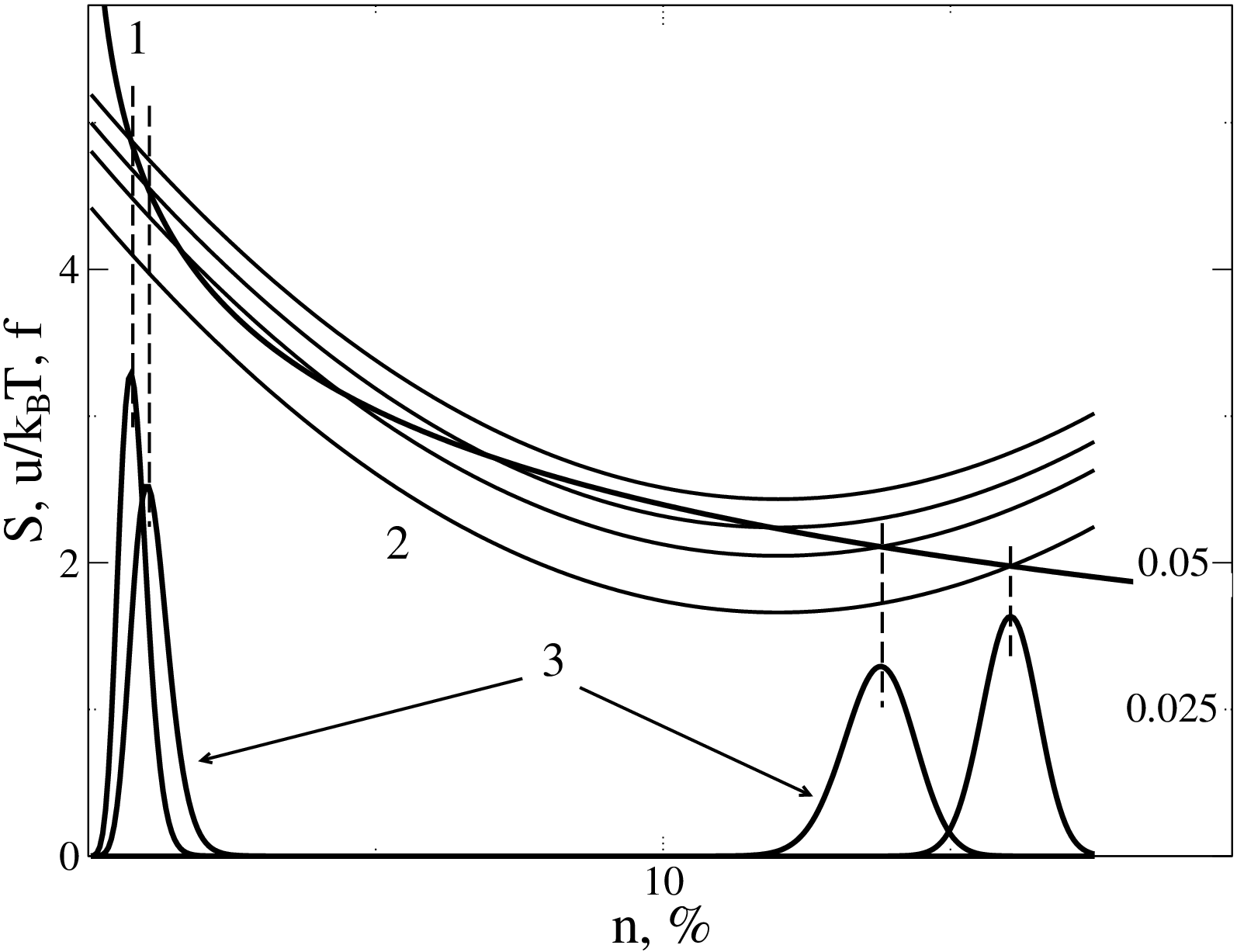}
\caption{\label{f3} Solution of Eq. \ref{b9} at quadratic dependence of energy of vacancy from the vacancy concentration. 
The fundamental curve  $S$ (1), energy of vacancy $u$ (2) and $PDF$ (3): $u_{0}: 0.135; 0.13; 0.125; 0.115 (eV)$. 
The rest parameters are $u_{1} = 0.6\cdot10^{-3} (eV)$, $u_{2} = 0.125\cdot10^{-5} (eV)$, $T=300K$. 
The right-hand scale is for PDF.}
\end{figure}

The first two terms in Eq. \ref{b9} are sums $S$ of the a slowly divergent harmonic series. This part of the equation 
depends only on system size and doesn't depend on material parameters. It is the fundamental part of Eq. \ref{b9} 
decreasing with the growing parameter $n$ (curve 1 of Fig. \ref{f1}, \ref{f2}, \ref{f3}). The last term in 
Eq. \ref{b9} depends on material parameters through coefficients $u_{k}$.

Now let us consider a problem in varying degree of approximation. In the linear approximation on the number of 
vacancies for the internal energy or at constant energy of vacancy there is only one solution of Eq. \ref{b9}, 
which coincides with the well known Frenkel solution \cite{f55} (see Fig. \ref{f1}). 
At lowering of the energy of vacancy or at 
the increase of temperature the solution is continuously displaced to region of higher number (or concentrations) 
of vacancies. We can formally get any value of vacancy concentration, but, it is clearly, that too high vacancy 
concentrations about $100$ percent lose physical sense and theory becomes useless. However Frenkel had proposed 
to suggest not too high vacancy concentrations about 10 percent, as transition to the liquid state. At the same time, 
the solution offered by him (in approximation of independence of energy of vacancy from the vacancy concentration) 
can not explain or describe the process of melting of solid, as saltatory phase transition of the first kind

In the quadratic approximation on the vacancy number for the internal energy or in the linear approximation 
for energy of vacancy Eq. \ref{b9} has already two solutions in a region of high energy of vacancy or in region 
of low temperatures (see Fig. \ref{f2}). 
One of them in region of lower concentration of vacancies, actually, coincides with 
the Frenkel solution, but it is some displaced due to interaction between vacancies. The second solution in region 
of higher concentration of vacancies describes the equilibrium or stationary solution of problem also; however, 
it corresponds to a minimum of PDF and is unsteady. Probability of realization of this state is less not only 
in relation to the stable stationary state but also in relation to any non-equilibrium state. 
In addition, for the low energy of vacancy or for high temperatures we get in a region in which the solutions 
of Eq. \ref{b9} are absent at all. The last circumstance testifies unsuitability of the approximation in this region, 
and it is need to resort to a higher approximation.

In the cubic approximation on the number of vacancies for internal energy or in the quadratic approximation 
for the energy of vacancy Eq. \ref{b9} can have one or three solutions already (see Fig \ref{f3}). 
In region of high values of the energy of vacancy or low temperatures equation there is one solution, 
which is the Frenkel solution modified due to the nonlinear contributions. At lowering of the energy of vacancy 
or with growth of temperature equation will have three solutions, one of which (left) is the modified Frenkel solution, 
second one (intermediate) is unsteady, and third one (right) in region of high concentration of vacancies can be 
possible to examine as proper the liquid state of a matter. In this case we have an equilibrium coexistence of 
solid and liquid phases of the matter. At the yet greater increase of temperature we pass to the region of 
existence of only one solution which corresponds to clear-melted matter. Thus a transition from the solid state 
to liquid one is carried out as a phase transition of the first kind.

In the article the deduction of a probability distribution function is considered for the case of general 
dependence of the internal energy from the number (concentration) of vacancies. Within the framework of 
such generalization it is shown the melting of solids can be described, as a saltatory phase transition of 
the first kind without bringing in of a concept of order parameter.

\begin{acknowledgments}
The work was supported by the budget topic № 0109U006004 of NAS of Ukraine.
\end{acknowledgments}

\end{document}